\documentstyle[twoside,12pt]{article}
\input{def}
\begin{document}
\catcode`@=11
\def\marginnote#1{}
%
\hyphenation{un-der-ground}
\hyphenation{tem-pe-ra-ture}
\hyphenation{dis-cri-mi-na-tor}

\begin{center}
{\Large Phase transition time delays in irradiated Superheated Superconducting
Granules}
\end{center}
\vspace{.5cm}
M.Abplanalp, C. Berger, G. Czapek, U. Diggelmann, M. Furlan, A.
Gabutti,
S. Janos, U. Moser, R. Pozzi, K. Pretzl, K. Schmiemann \\
{\it Laboratory for High Energy Physics, University of Bern,
Sidlerstrasse 5, CH 3012 Bern, Switzerland} \\
 \\
D. Perret-Gallix \\
{\it LAPP, Chemin de Bellevue, 74941 Annecy, France} \\
 \\
B. van den Brandt, J.A. Konter, S. Mango \\
{\it Paul Scherrer Institute, CH-5232 Villigen PSI, Switzerland} \\
 \\
\begin{center}
\section*{Abstract}
\end{center}
\textwidth=5in
\baselineskip=12pt
\small
The time difference between a particle interaction in a Superheated
Superconducting Granule (SSG) and the resulting phase transition signal has
been explored. Detectors containing Zn and Sn SSG were irradiated with neutrons
and protons to study the heating mechanism taking place in nuclear recoil and
ionizing events. Scattered neutrons have been detected by a scintillator
hodoscope behind the SSG with a recoil energy measurement resolution of 10\%
and an interaction time resolution of 1ns. The fast transition of the
metastable granules allowed to determine the elapsed time between an energy
deposition and the phase transition signal. In the case of Sn granules,
the results show that the time
distributions are narrow and independent of the deposited energy in nuclear
recoil and ionizing events.
In Zn, however, the time distributions are much broader and depend
on the energy deposition in the granule. \\
 \\
\textwidth=6in
\baselineskip=14pt
\normalsize
\section{Introduction}
Superheated Superconducting Granule (SSG) detectors are being presently
developed for dark matter and neutrino detection as well as for x-ray
astronomy. They consist of a collection of tiny spheres embedded
in a dielectric medium in a magnetic field, kept at constant temperature in
the metastable state. The detection principle is based on the phase transition
from the metastable to the normal conducting state due to the energy
deposited by the interacting particle.
Various groups have already proven the ability of SSG to detect minimum
ionizing particles, x-rays and nuclear recoils.
It has been shown that the sensitivity of SSG to deposited energies of some
MeV depends on the interaction point and on the energy transport
properties inside the granule. An overview of the SSG detector
development is given in Ref.$^1$. \\
The time sequence of the physical processes which take place after a particle
interaction is
sketched in Fig.~1. The temperature inside the granule changes with time,
leading to a penetration of the magnetic field into the granule. The
resulting change of magnetic flux induces a voltage signal in a
pickup coil wound around the detector.
A trigger in the readout electronics then defines the time of
the signal occurrence. A measurement of the time difference between the
particle
interaction and the output signal then reveals information about the energy
transport phenomena and the dynamics of the field
penetration in the granule. The heating duration shifts the detector signals,
while the phase transition duration influences also the signal shape
of the readout
circuit which reacts on the magnetic flux change. A simultaneous analysis of
the time delay and the shape of the detector signals therefore determines both
the heating and the phase transition time of irradiated SSG. \\
  Protons deposit energy by ionization across
their way through the granules while neutrons interact with SSG by pointlike
nuclear elastic scattering.  The aim of the experiment was to explore the
dynamics of the heating processes and their duration in dependance on the
detector properties and the interacting particle type.
\section{Experiment}
To study the time differences, detectors containing Sn and Zn SSG were
irradiated with neutrons and protons at the Paul
Scherrer Institute (PSI), Switzerland. The neutron beam consisted of 70MeV
neutrons produced by irradiating a Be target with 72MeV protons coming from
the Injector I of PSI. Protons with about 50MeV energy have been produced by
irradiating a CH$_2$ target with the neutron beam. The SSG detectors with
0.1cm$^3$ volume consisted of about 10$^6$ granules embedded in plasticine
with a volume filling factor of about 10\%. The diameter distributions of
the SSG were flat between 15$\mu m$ and 20$\mu m$ for Sn and 28$\mu m$
and 30$\mu m$ for Zn SSG. Signals have been detected by pickup coils of
about 200 windings around the detectors, connected to J-FET amplifiers working
at room temperature. The signals were LCR-oscillations of the coils
together with the capacitance and resistance of the readout circuit with a
1.5$\mu s$ period, two typical signals of the Zn SSG are shown in Fig.~2.
The detectors were
exposed to a magnetic field produced by a Helmholtz magnet outside the
cryostat.
Preparation of the detectors has been performed by rising the field to a value
B$_1$ and then reducing  it to a slightly lower value B$_2$ The detector
threshold is then defined as h=(B$_1$-B$_2$)/B$_1$ and was varied between 0.5\%
and 5\% for the Sn and between 1\% and 10\% for the Zn SSG detectors. \\
In the case of neutron irradiation, the recoil interaction has been triggered
by a scintillator hodoscope 200cm behind the SSG consisting of 18 rods of
5*5cm$^2$ cross section and 150cm length, detecting neutrons scattered by the
SSG$^2$. The hodoscope was read out by 36 photomultipliers, allowing time
measurements with about 1ns resolution. The position of the interacting
neutron inside a single element was determined by measuring the time difference
between the signals of two photomultipliers located at the ends of the bar
with a resulting spatial resolution of $\pm$4cm corresponding to an angular
resolution of $\pm$20mrad. Therefore, the energy deposited to the SSG can be
evaluated with about 10\% resolution by
\begin{equation}
E_r = 4 sin^2(\vartheta / 2) \frac{E_n}{m_{nuc}}
\end{equation}
with E$_r$ the recoil energy, $\vartheta$ the scattering angle, E$_n$
the kinetic energy of the neutron and m$_{nuc}$ the nuclear mass.
Energies between one and 300keV have been observed. \\
During irradiation with protons, the interaction was triggered by
a 3mm thick scintillator placed in the beam in front of the SSG, allowing time
measurements with 1ns resolution.  \\
The SSG signal trigger was set at the time of the first zero
crossing of the signal oscillation. The advantage over a discriminator
threshold is its independence of signal height.\\
The time between the interaction and the signal trigger is offset from the time
difference between the real interaction and the
SSG signal by a constant value due to
delays in the experimental electronics. Although an absolute timing measurement
could not be obtained, dependances of the time differences on the detector
threshold, the particle type and the deposited energy within a SSG have been
studied.
\section{Experimental results}
The time distributions of the Sn SSG detector at two different detector
thresholds (h=1\% and h=5\%), when irradiated with neutrons and
protons, are shown in Fig.~3.
The distributions for the protons are very narrow
with a FWHM of about 40ns. The time resolution of the clock used for the
zero crossing trigger was 20ns. When irradiated with neutrons, the SSG
respond with a broader time spectrum (FWHM$\simeq$100ns).
The energy loss due to continuous ionization of a proton with 50MeV kinetic
energy across a Sn granule with 15-20$\mu$m diameter was derived by the GEANT
program$^3$ to be about 150-300keV.
The results indicate that the heating process in Sn SSG is very fast and
independent on the detector threshold if the interacting particle deposits
energy of some hundred keV by ionization along a track through the granule,
whereas in pointlike
nuclear recoils (1-100keV) the heating process takes place within a time
scale of about 100ns for a granule with 15-20$\mu$m diameter. \\
The time distributions of the Zn detector at a threshold of h=2\% for three
different recoil energies and for the proton irradiation are shown in Fig.~4.
The width varies from 300ns FWHM for the protons (energy losses calculated by
GEANT are about 150-350keV) to 800ns FWHM for small recoil energies. Two
signals taking place at different times are shown in Fig.~2. Besides their
different time delays the pulse shapes do not differ, indicating that in
both events
the phase transition time is the same and that the width of the time
distributions is dominated by the spread of individual heating times.\\
Fig. 4 shows that
the time difference decreases with increasing recoil energy. The peak of the
distribution for 5-15keV is located at 1.9$\mu$s, for 150-300keV it is at
1.5$\mu$s. High recoil energies drive the granule faster into the
normal phase.
\section{Conclusion}
The relative time delay between a particle interaction and the phase transition
signal in Sn and Zn SSG detectors has been studied for ionizing and nuclear
recoil events. The width of the delay distributions were narrower in ionizing
than in nuclear recoil events. The signal shapes turned out to be the same in
both cases, indicating that the observed time delays are primarily due to the
heat diffusion mechanism and not to fluctuations in the magnetic field
penetration time within the granules. The heating time scale for the Sn SSG
(diameter 15-20$\mu$m) turned out to be about 100ns and for the Zn SSG
(diameter 28-30$\mu$m) about 1$\mu$s. \\
It was found that for higher deposited energies inside the granules the heating
time fluctuations become smaller. The observed large difference in the heating
time scale between Sn and Zn SSG is a reflection of the large difference of the
quasiparticle lifetime in the two materials, leading also to the observed local
and global heating effects$^4$. \\
This work was supported by the Schweizerischer Nationalfonds zur Foerderung der
wissenschaftlichen Forschung and by the Bernische Stiftung zur Foerderung der
wissenschaftlichen Forschung an den Universitaet Bern.

\section*{Figure Captions}
\begin{itemize}
\item [Fig. 1] Time sequence of some properties. a) Interaction trigger, b)
Temperature at the point where the magnetic field starts to penetrate, c)
Magnetic flux through the pickup coil around the granule, d) Induced voltage by
the readout circuit, e) First zero crossing trigger of the readout
signal.
\item [Fig. 2] Two signals of the readout electronics from the Zn
SSG detector.
\item [Fig. 3] Time difference distributions for the Sn detector. The time
scale
is offset by a constant value due to electronic delay.
\item [Fig. 4] Time difference distributions for the Zn detector at threshold
h=2\%, for three recoil energies and for proton irradiation. The time scale
is offset by a constant value due to electronic delay.
\end{itemize}


\begin{thebibliography}{99}
\bibitem{pretzl1} K. Pretzl, Particle World {\bf 1/6} (1990), 153 and J. Low
Temp. Phys. {\bf 93} (1993), 439.
\bibitem{recoil} C. Berger et al., Nucl. Instr. Meth. {\bf A330} (1993), 285
and M. Abplanalp et al., J. Low Temp. Phys. {\bf 93} (1993), 491
\bibitem{geant} R. Brun et al., GEANT3 User's guide, CERN DD/EE/84-1
\bibitem{frank} M. Frank et al., Nucl. Instr. Meth. {\bf A287} (1990), 583
\end{thebibliography}
\end{document}